\documentclass{article}

\usepackage{graphicx,epsfig}

\begin{document}
\Large
\begin{center}
\textbf{Effects of external global noise on the catalytic CO oxidation on Pt(110)}

\vspace{0.5cm}

\normalsize
P. S. Bodega $^{1}$, S. Alonso $^{1,2}$, H. H. Rotermund $^{3,4}$

\end{center}
\small
1.	Fritz-Haber Institute of the Max Planck Society, Faradayweg 4-6, 14195 Berlin, Germany\\
2.	Physikalisch-Technische Bundesanstalt, Abbestrasse 2-12, 10587 Berlin, Germany\\
3.	Department of Physics and Atmospheric Science, Dalhousie University, Halifax, NS B3H 3J5 Canada\\
4.	Corresponding author,  e-mail: harm.rotermund@dal.ca\\

\normalsize

Oxidation reaction of CO on a single platinum crystal is a
reaction-diffusion system that may exhibit bistable, excitable, and
oscillatory behavior. We studied the effect of a stochastic signal
artificially introduced into the system through the partial pressure of
CO. First, the external signal is employed as a turbulence suppression
tool, and second, it modifies the boundaries in the bistable transition
between the CO and oxygen covered phases. Experiments using photoemission
electron microscopy (PEEM) together with numerical simulations performed
with the Krischer-Eiswirth-Ertl (KEE) model are presented.

%===================================================================\section{Introduction}
\section{Introduction}
\label{sect:intro}

Chemical reactions in extended media are one of the characteristic systems where
pattern formation outside equilibrium can be observed. The nonlinearity of such reactions and the
diffusion of its components induce the formation of complex patterns. The Belousov-Zhabotinsky
reaction and the catalytic oxidation of CO on single crystal platinum surfaces are  chemical
examples where spiral waves, target patterns, or chemical turbulence have been studied \cite{kapral-showalter-book}. Such
systems have been extensively employed because it is relatively easy to control the chemical and physical conditions
which are responsible for the formation of the spatiotemporal structures \cite{mikhailov_06}.

Natural systems are unavoidably subject to random fluctuations (noise), stemming from either environmental variability or thermal effects. These undesirable perturbations mask and affect
the deterministic dynamics of the system. Such fluctuations gain importance
far from equilibrium \cite{haken_book_85,nicolis_book_77} because they can interact with
the nonlinearities of the system, giving rise to a rich variety of behaviours \cite{horsthemke_84}.
The effects
of noise in extended active media \cite{sagues_07} are particularly interesting. Chemical reactions are one of the
most convenient systems used to study this type of phenomenology. Both Belousov-Zhabotinsky
and chlorine dioxide-iodine-malonic acid (CDIMA) reactions are good examples where the effects of noise
have been studied extensively.
The photosensitive version of such reactions
is particularly suitable for the introduction of controlled external fluctuations through a
stochastic illumination \cite{kadar_98,n-alonso-sendina-2001,alonso_08}, where the characteristic time and length
can be easily tuned by a computer. The most significant results have been obtained for noises
with spatial structure. However, interesting results have also been observed with homogeneous noisy illumination
\cite{n-sandina_2000} or applying an stochastic electric field \cite{zhou_02}.

Heterogeneous catalytic CO oxidation is well known for the wealth of variety of pattern formations that can develop on the surface of a single crystal. Depending on the partial
pressure of the species inside the reactor chamber, different types of patterns have been observed: spiral waves, target patterns, standing waves, and turbulence \cite{jakubith-1990}.
This reaction is not photosensitive, but the pressure of the reactants
can be externally controlled \cite{bodega_07}; in particular, a stochastic signal can be artificially introduced.
There is no spatial structure in the external signal and therefore the noise is homogeneous.
Previous experimental and analytical studies in bistable reactions show how noise can anticipate the transitions
between both bistable states in CO oxidation in Ir(111) \cite{wehner_05,hoffmann_06}, and how local fluctuations become important 
\cite{pineda_06}. Theoretical predictions of stochastic resonance in the catalytic oxidation of CO in Pt(110) have also been reported
\cite{yang_98}.

The effects of external noise can be interpreted as a control mechanism of the
different types of patterns, and by tuning the characteristics of these external
fluctuations, transitions and new states can be achieved.
The control of chemical turbulence is a problem tackled from a number of
different angles \cite{mikhailov_06,schuster_book}. In the CO-Pt system for instance,
attempts have been made with feedback techniques
\cite{bertram-beta-2003a,beta-bertram-2003}, local control
\cite{stich_punckt_beta_08}, periodical global forcing
\cite{bertram-beta-2003b}, and resonant global forcing \cite{bodega_07}.

Two different scenarios of noise effects are considered here:
first, we study the elimination of
turbulence through stochastic forcing, and second, we focus on the  
noise related modification of the boundaries in the bistable transition between the CO and oxygen covered phases.
Both experimental studies are contrasted with the results of numerical simulations performed with
the KEE model for the CO oxidation in Pt(110) \cite{krischer-eiswirth-ertl-1992a}. Finally, we discuss
the results and compare them with previous control strategies for the catalytic CO oxidation.

%===================================================================\section{Experimental Setup}
\section{Experimental setup}
\label{sect:exp}
\begin{figure}[]
\begin{center}
\includegraphics[width=13cm]{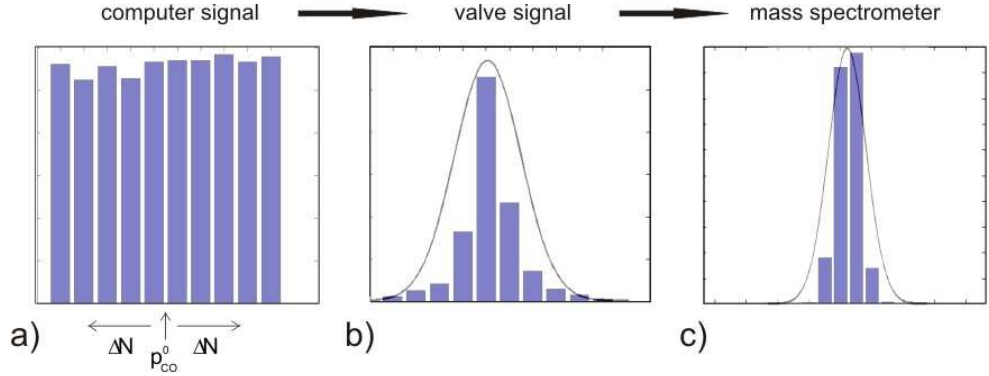}
\end{center}
\caption{(Color online) Histograms of the $p_{CO}(t)$. a) Generated signal, b) signal in the dosing system, and c) mass spectrometer signal inside the reaction chamber (Gaussian distribution overplotted).}
\label{f:noise_integration}
\end{figure}

To perform the experiments, a single crystal Pt(110) is mounted inside a constant flux ultra high vacuum (UHV) chamber, where CO and oxygen are constantly added into,
while the total pressure is kept constant by continuously pumping the entire system. A PEEM \cite{rotermund-1992} allows us to spatially 
resolve the local work function of the adsorbate species and to
record it with a CCD camera at a rate of 25 full frames per second. A meaningful modulation of the CO partial pressure was introduced into the system as a computer controlled source of random fluctuations following:
\begin{equation}
   p_{CO}(t_{i}) = p^{0}_{CO} (1 + \psi_{i} \cdot \bigtriangleup N ) \label{eq:COnoise}
\end{equation}
where $\psi_i$ is a random number between -1 and +1 that is regenerated by the computer in time intervals of $\bigtriangleup t$ ($t_{i}-t_{i-1}=\bigtriangleup t$). $\bigtriangleup N$ is the percentage of CO partial pressure that is added to the base value. The combination of the parameters $\bigtriangleup t$ and $\bigtriangleup N$ determines how fast and how much $p_{CO}(t)$ changes, or in other words, the magnitude of the multiplicative noise present in the reaction. The two parameters are then used by the computer to control the CO gas dosing valve. Due to the solenoid valve effective reaction time, and the low pass filtering
between the dosing system and the
chamber, the generated rectangular signal is integrated by the system, resulting in a Gaussian-like distribution in the solenoid valve
and in the reaction chamber, as shown in Fig. \ref{f:noise_integration}. Hence, the CO partial pressure in the reaction environment has a fluctuating value between
$p^{0}_{CO}+\triangle N$ and $p^{0}_{CO}-\triangle N$, homogeneous in space, characterized by a histogram like the one depicted in Fig. \ref{f:noise_integration} b which
resembles a Gaussian noise.
\\

%===================================================================\section{Theoretical Model}
\section{Reaction model}
\label{sect:theorie}

In this work we use the KEE model \cite{krischer-eiswirth-ertl-1992a}, which has been previously employed to 
study CO oxidation in surfaces of Pt(110) \cite{bertram-mikhailov-2003,Wolff-2004}. The 
model consists of three coupled equations for the CO and oxygen coverage,
respectively $u$ and $v$, and the fraction of the surface found in the nonreconstructed Pt $1
\times 1$ phase, corresponding to the third variable $w$. The equations read:
\begin{eqnarray}
\partial_t u &=&D \nabla^2 u + k_1 s_{CO} p_{CO}  (1-u^3) - k_2 u - k_3 uv , \label{keeu} \\
\partial_t v &=& k_4 p_{O_2} [ s_{O,1\times1} w + s_{O,1\times2} (1-w) ] (1-u-v)^2  - k_3 uv, \label{keev} \\
\partial_t w &=& k_5 \left(  \frac{1}{1+exp \left( \frac{u_0 -u}{\delta u} \right)}
- w \right). \label{keew}
\end{eqnarray}

For a brief explanation of the model and the parameter values used here, see Table I. Such parameters
correspond to previous numerical studies of the same experimental system under similar conditions
\cite{bertram-mikhailov-2003}. The values of the partial pressures of CO and oxygen are respectively
$p_{CO}$ and $p_{O_2}$, and they are specified for each case studied here. Several variants of this model have been previously employed 
to study different types of bifurcations observed in the CO oxidation or the dynamics
of spiral waves \cite{baer-1993}. The effect of the temperature can also be considered, assuming an Arrhenius-type dependence on the temperature of the model parameters 
\cite{Wolff-2004}. Here, we keep the parameter corresponding to the temperature constant $T=542.35$ $K$.

\begin{figure}[h!]
\begin{center}
\setlength{\unitlength}{1mm}
\includegraphics[width=6.5cm]{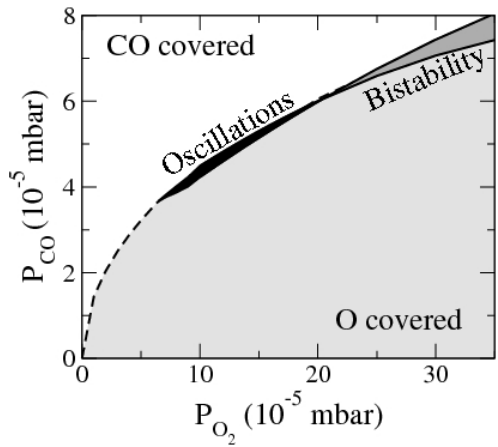}
\end{center}
\caption{Phase diagram of the KEE model keeping the values of table I constant. This diagram does not take into account the diffusion. Regions of oscillations and bistability
are marked respectively as black and dark gray areas.}
\label{f:phase_diagram}
\end{figure}

The phase diagram of the KEE model for both partial pressures ($p_{CO}$ and $p_{O_2}$), keeping the rest of the parameters constant, is shown in Fig. \ref{f:phase_diagram}. 
It corresponds to the system without diffusion, thus there is no information about spatial patterns. As expected for large values of the pressure of oxygen and low pressure of
CO, the oxygen covered phase is stable. On the other hand, for large values of the pressure of CO and low pressure of oxygen, the CO covered phase is also stable. For intermediate values
of both partial pressures (typically with the pressure of oxygen three times larger than the pressure of CO), both phases compete. While for large pressures this competition gives rise to bistability and
hysteresis, for intermediate pressures it also induces oscillations. In this region of periodic alternation, not only global oscillations but turbulence are observed depending on the
values of the parameters. For low pressures of both CO and oxygen, a continuous transition between both phases is observed (dashed line in Fig. \ref{f:phase_diagram}. This diagram qualitatively 
agrees with the experiments of CO oxidation on Pt(110).

In order to mimic the experimental conditions of random external modulations of the partial pressure of CO, we add a stochastic variable to the value of the pressure:

\begin{eqnarray}
p_{CO} = p^0_{CO} \left( 1 + \xi_e(t) \right)
\end{eqnarray}
where the variable $\xi_e(t)$ is a Gaussian spatially homogeneous white noise with null mean $ <\xi_e(t)> = 0$ and variance $<\xi_e(t)\xi_e(t')> = 2 \sigma^2 \delta(t-t')$, where
$\sigma^2$ is the intensity of the noise. 

It is known that the surface of the crystal is not completely homogeneous and there are heterogeneities which affect the propagation of the surface waves qualitatively. 
Such heterogeneities are consequence of several experimental factors (facetting, subsurface species or cristal surface defects) and their activity may depend on several processes
like the local concentration of CO and O in an unknown way.
In order to consider these internal fluctuations, we introduce an additive noise term $ \xi_i(x,t)$ in the first equation of the model Eq.(\ref{keeu}).
This variable is Gaussian white noise, distributed in space with null mean $ <\xi_i(x,t)> = 0$ and variance $< \xi_i(x,t)\xi_i(x',t')> = 2 \epsilon_i \delta(x,x') \delta(t-t')$, where
$\epsilon_i$ is the intensity of the noise. We note that these internal fluctuations are not related with the external experimental
noise which has been already introduced above as multiplicative noise. Since we are not going to do a systematic study of these internal fluctuations, we are keeping the 
intensity of this noise low and constant ($\epsilon_i=0.0005$) for all of the numerical simulations following.
They will act as precursors in the simulations during transitions.

In summary, we introduce two types of random processes: first, a homogeneous random noise through the parameter $p_{CO}$, mimicking the external experimental noise introduced 
by the CO pressure, which is the main object of study in this manuscript. Second, a spatiotemporal noise, that takes into account different experimental sources of randomness. 

We integrate Eqs. \ref{keeu}-\ref{keew} using a finite differences scheme and the Heun method \cite{sancho_book} for
stochastic partial differential equations. Two-dimensional
numerical simulations are completed using a finite differences scheme with a spatial and temporal discretization of  $\delta x = 2 \mu m$ and $\delta t = 0.001 s$ respectively. We
employ periodic boundary conditions in all of the simulations.

\begin{table}[h!]
\caption{Parameters of the reaction model.}
\bigskip
%\begin{tabular}{|c|c|c|}
\begin{tabular}{ccc}
\hline
Parameter& Value & Meaning\\
\hline \hline\\

$k_1$   &   $3.14 \times 10^5 mbar^{-1}$  & Impingement rate of CO \\
$k_2$   &   $10.21 s^{-1}$  & CO desorption  rate \\
$k_3$   &   $283.8 s^{-1}$  & CO Reaction  rate \\
$k_4$   &   $5.86 \times 10^5 s^{-1} mbar^{-1}$  & Impingement rate of O$_2$ \\
$k_5$   &   $1.61 s^{-1}$  & Phase transition  rate \\
$s_{CO}$   &   $1.0$  & CO sticking coefficient \\
$s_{O_{1 \times 1}}$   &   $0.6$  & Oxygen sticking coefficient on the $1 \times 1$ phase \\
$s_{O_{1 \times 2}}$   &   $0.4$  & Oxygen sticking coefficient on the $1 \times 2$ phase \\
$u_0$, $\delta u$   &   $0.35$, $0.05$  & Parameters for the structural phase
transition \\
$D$ & $40 \mu m^2 s^{-1}$  & CO diffusion coefficient \\
$u_{ref}$ & 0.3358  & Reference CO coverage  \\
\hline \hline
\end{tabular}
\end{table}

\section{Turbulence suppressed by noise}
\label{subsect:noise_turb}

Spatiotemporal chemical turbulence spontaneously develops on a platinum crystal surface for a window of values of the reaction parameters (reactants, partial pressures, and
temperature). Such
turbulence is characterized by the continuous creation and annihilation of phase defects on the surface \cite{beta_06}. Although it corresponds to a complex state, a 
characteristic frequency
can still be obtained. Such turbulence can be effectively suppressed by a noise modulation of the reactants, namely $p_{CO}(t)$. 

\begin{figure}[h!]
\begin{center}
\includegraphics[width=11cm]{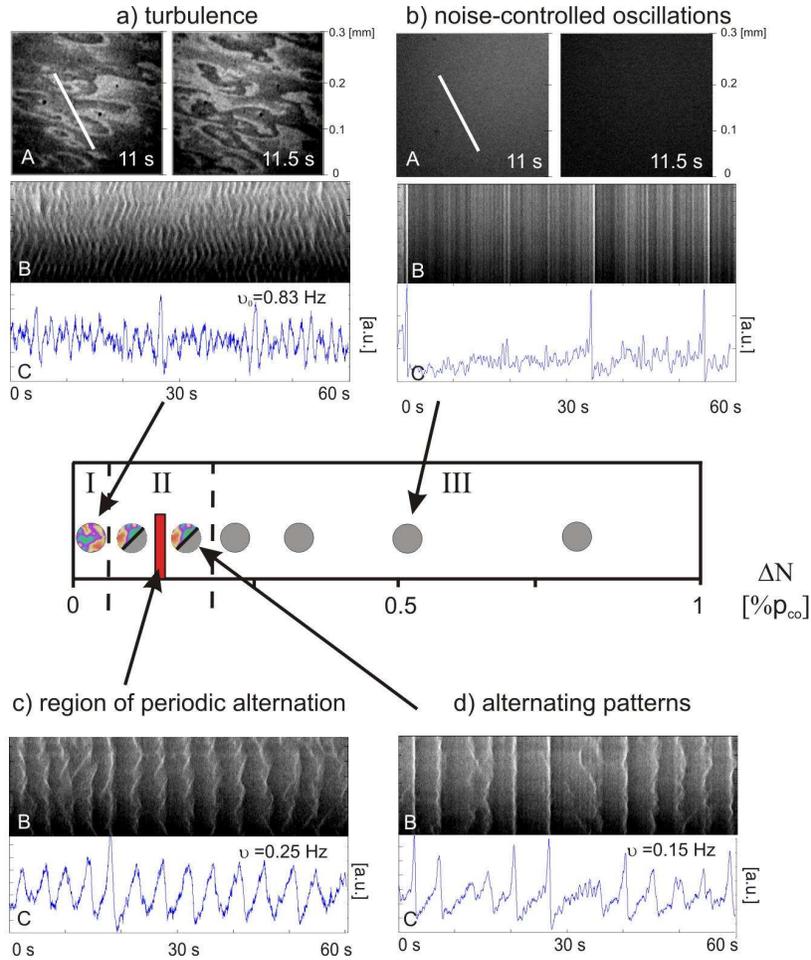}
\end{center}
\caption{(Color online) PEEM snapshots A of the Platinum surface and space-time plots B, for different
values of $\Delta N$ applied on an initial turbulence state of the chemical reaction. The spacetime
plots represent the reaction pattern development in time along the white bar depicted on A. The
corresponding pixel image intensities are plotted on C. Reaction parameters are T = $540K$, 
$p_{O_2}$ = $1.3\cdot 10^{-4} mbar$, 
$p^0_{CO}$ = $6.6\cdot 10^{-5} mbar$, and $\Delta t$ =0.5 s. Three regime differentiated
areas are pointed out (I, II, and III), showing turbulence (a), noise-controlled oscillations (b), periodic alternation (c), and alternating patterns (d), 
which are further commented on the text.}
\label{f:noise_suppression}
\end{figure}

Fig.
\ref{f:noise_suppression} represents the corresponding spatiotemporal patterns
seen when different noise intensities are applied to a turbulent regime. In the absence 
of external noise, spiral
turbulence develops freely (see 2D images in Fig.(\ref{f:noise_suppression},a)). For large
intensities of the fluctuations (above 0.2 \% $p_{co}$), the noise completely annihilates the reaction
dynamics; the whole system then follows the noisy signal Fig. \ref{f:noise_suppression}b. For intermediate 
noise intensity, the system is driven to a partially suppressed state featuring an alternation between asynchronized 
and homogeneous global oscillations Fig. \ref{f:noise_suppression}d. It
corresponds to a competition between the turbulence behavior of the surface and the global
synchronization produced by the noise.
This alternation was observed to be regular and periodic for a particular range of parameters Fig. \ref{f:noise_suppression}c.
 Remarkably, the frequency of such responses are different than the characteristic frequency of the spiral turbulence.
Finally, noise reaches values that completely annihilate the reaction dynamics; the system then follows the noisy signal.

The natural frequency of the system is affected by the noise. Fig. \ref{f:noise_suppression}
shows how the periodicity of the local oscillations changes with $\Delta N$ visible within the snapshots and the space-time plots, and compares the
initial spiral turbulence of region I with the
periodic induced patterns within region II, with the noise controlled region
III for a particular fixed value of $\Delta t$.
Region III has no local periodicity whatsoever and reflexes the imposed random
oscillations.

Region II of Fig. \ref{f:noise_suppression} encompasses the range of noise parameters in which the original spiral turbulence is not completely suppressed. Interaction with noise
drives the active media through transient states, alternating between spatiotemporal chaos and a lack of patterns. Furthermore, for certain parameters these cycles  may become
regular. At the bottom left of Fig. \ref{f:noise_suppression} suppressed patterns alternating with re-activated turbulence in
regular cycles of 4 seconds can be seen. The noise parameters corresponding to this induced behavior draw a narrow stripe in the turbulence suppression map (seen red bar in Fig.
\ref{f:noise_suppression}). Along that stripe, the interplay between excitability, refractory time, noise, and diffusion results in time regularity.

Larger or faster modulations than those showed in Fig. \ref{f:noise_suppression} are not considered here because of the limitations of the experimental setup (experiments showed strong
disagreement between the computer values and the real state in the chamber for values $\Delta t <$ 0.1 or $\Delta N >$ 3 \% $p^{0}_{CO}$). Therefore this study
project was specifically focused on the smallest fluctuations with a relevant impact.

\begin{figure}[h!]
\begin{center}
\includegraphics[width=12cm]{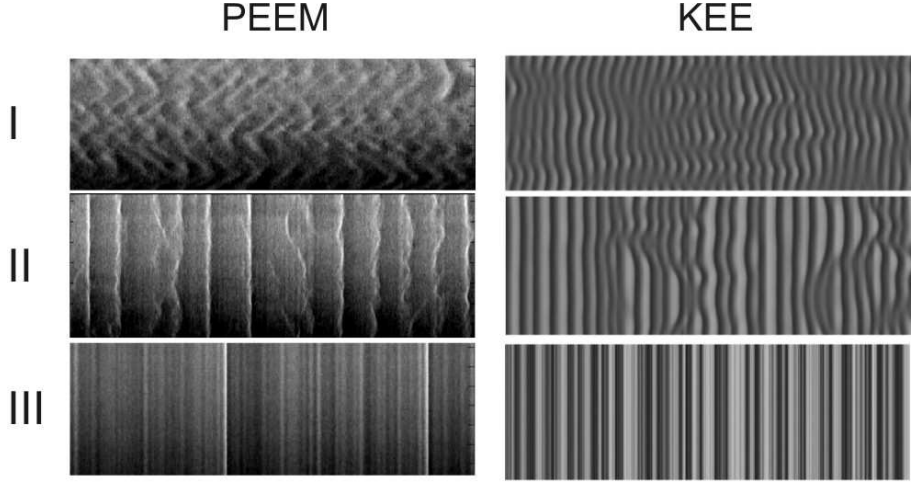}
\end{center}
\caption{Compared space-time plots corresponding to the turbulence suppression map regions I, II, and III; experiments (PEEM) vs. simulations (KEE). 
Simulations are performed with the
values of the partial pressures: $p_{O_{2}} = 1.3 \cdot 10^{-4}$ mbar and $p^0_{CO} = 4 \cdot 10^{-5}$ mbar in a system of size 
$0.2 \times 0.2$ $mm^2$. From top to botton the corresponding noise intensities are 
$\sigma$= 0.0 \% $p^0_{CO}$, $\sigma$= 0.25 \% $p^0_{CO}$ and $\sigma$= 2.5 \% $p^0_{CO}$.}
\label{f:XT_comparacionKleine}
\end{figure}

The series of space-time plots depicted in Fig. \ref{f:XT_comparacionKleine} show the typical surface dynamics of the three regions (I, II, and III) adressed in Fig.
\ref{f:noise_suppression} compared with numerical simulations; one can see that the plots based on the KEE model qualitatively agree with the experimental
results. Wave turbulence (I), alternation of global oscillations and wave turbulence (II), and dominance of the noise (III) are observed in both cases.

The results of the numerical simulations agree and confirm the experimental results. Under parameter values corresponding to turbulence in the KEE model, the introduction of a
global stochastic signal may induce the global coordinating of the whole system. For intermediate values of the external signal a competition of the global coupling and the deterministic
turbulent dynamics is observed, giving rise to intermittent turbulence and a number of intermediate patterns. For some noise intensities, periodic oscillations are observed and even standing waves or pacemakers are
temporarily created by the global noise.

In the numerical simulations, the frequency of the spatiotemporal patterns also
decreases with the noise intensity. The patterns of the chemical turbulence can
be characterized by large frequencies. The global noise not only couples the
whole system but also reduces the frequency of the patterns.

%===================================================================\subsection{Bistable Transition}
\section{Transitions anticipated by noise}
\label{subsect:noise_trans}

\begin{figure}[t!]
\begin{center}
\includegraphics[width=12cm]{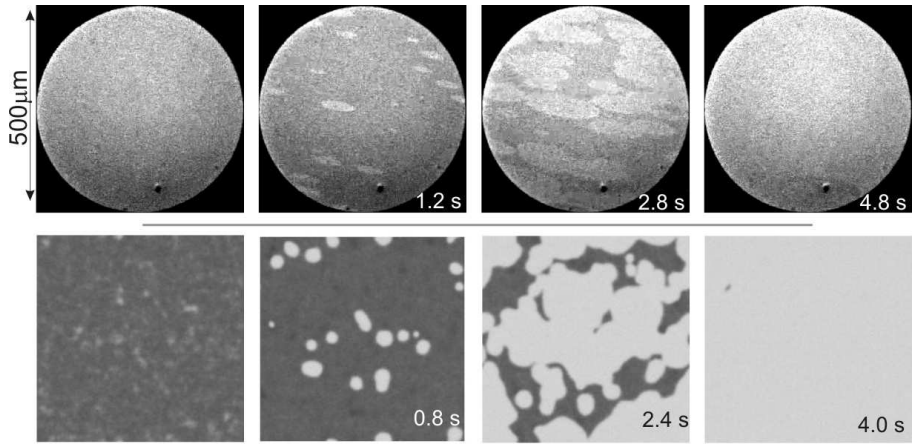}
\end{center}
\caption{PEEM snapshots (top) compared with numerically generated images (bottom). Different stages of the bistable transition, oxygen covered 
to CO covered can be seen.
Experimental parameters are: T = 509 K, $p_{O_{2}} = 1.1 \cdot 10^{-4}$ mbar, $\Delta t$= 0.25 s, and $\Delta N$ = 0.5  
\% $p^{0}_{CO}$. Parameters used for the
simulation are: $p_{O_2} = 4.5 \cdot 10^{-4}$, 
 $p_{CO} = 9.08 \cdot 10^{-5}$ in a system of size 
$0.4 \times 0.4$ $mm^2$, and internal noise $\epsilon_i=0.0005$.}
\label{f:bubbling}
\end{figure}

Further experiments were focused on the bistable
transition from a mainly oxygen covered Pt crystal to a mainly CO covered
state. These experiments start with a certain amount of oxygen in the reaction
chamber ($p_{O_{2}} = 1.1 \cdot 10^{-4}$~mbar) with no CO. Once the
surface is oxygen covered, the $p_{O_{2}}$ is kept constant and then the $p_{CO}$
is increased while adding additional fluctuations to it. For a large enough CO partial pressure, all of the
adsorbed oxygen reacts away releasing CO$_{2}$, the sample is completely CO
covered and the transition is complete. This process is clearly discernible in the PEEM images
(top of Fig. \ref{f:bubbling}): it is seen as the change from a uniform dark area to a uniformly bright one through a number of growing
bubbles. Due to the introduction of the additive noise in Eq. \ref{keeu}, the KEE model casts similar transition patterns (bottom of Fig. \ref{f:bubbling}). The intensity of such additive noise has been tuned to qualitatively reproduce the experimental patterns shown in Fig.
\ref{f:bubbling}, and kept constant for all of the simulations. Without such noise the system basically behaves homogeneously and 
no bubbles would be observed.

\begin{figure}[h!]
\begin{center}
\includegraphics[width=10cm]{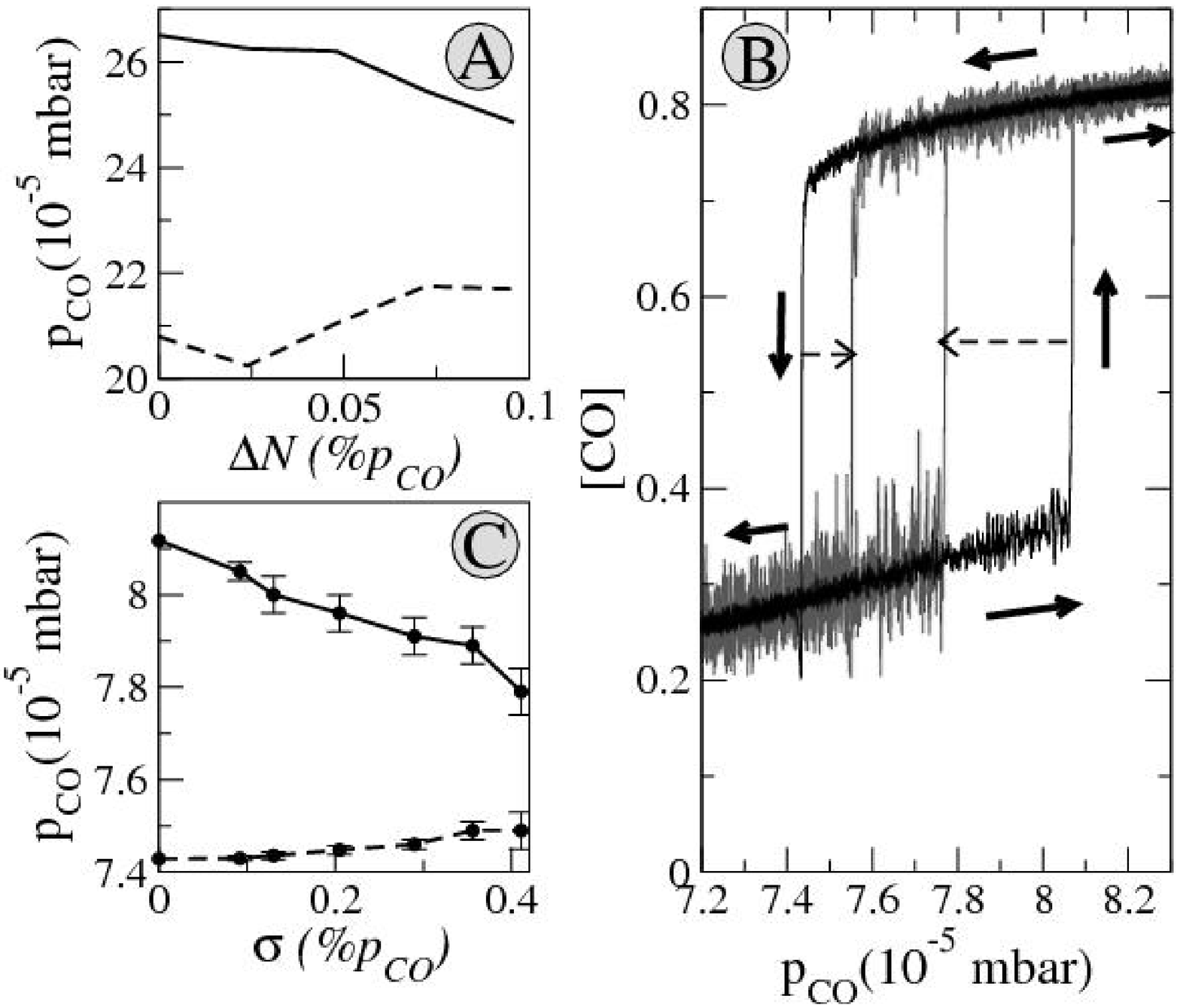}
\end{center}
\caption{Panel a) shows the critical $p_{CO}$ of both transitions, from oxygen covered to CO covered (solid line) and from CO covered to oxygen covered platinum (dashed line),
for a value of applied noise. Reaction parameters are: T = 509 K,
$p_{O_{2}} = 1.1 \cdot 10^{-4}$ mbar, and $\Delta t =0.1$ s. Panel b) shows the hysteresis loop for two noise intensities $\sigma$= 0.09 \% $p_{CO}$ and  $\sigma$= 0.28 \% $p_{CO}$. 
Panel c) represents the corresponding numerical simulation plotted in terms of the simulation
parameter $\sigma$. Numerical simulations have been completed with $p_{O_{2}} =
3.5 \cdot 10^{-4}$ mbar in a system of $0.2 \times 0.2$ $mm^2$.}
\label{f:transitions}
\end{figure}

The same steps are followed backwards for the back-transition to the initial CO covered
state. Nevertheless, the critical $p_{CO}$ is lower in this case, which is
explained by the asymmetric inhibition of the reaction: the oxygen covered
Pt does not behave in the same way as the poisoned CO covered surface
\cite{imbihl-ertl-1995}. There is an inherent hysteresis loop in this bistable
transition due to the asymmetry of the reaction. Fig. \ref{f:transitions}A describes 
how the mentioned hysteresis loop shrinks as the noise applied is larger.


Equivalent numerical simulations in the bistable region reproduce the main features of the hysteresis loop observed in the experiments. Starting from an oxygen covered phase, the
pressure of CO is slowly increased inducing the transition to the CO covered phase. The posterior reduction of the CO pressure shows a lower value for the transition, giving rise to
the hysteresis loop. Two numerical realizations with different noise intensities of this loop are shown together in 
Fig.~\ref{f:transitions}B. Larger noise intensities imply smaller hysteresis loops. This reduction of the size of the loop is plotted in Fig.~\ref{f:transitions}C.

\section{Discussion}
\label{sect:noise_disc}

When the CO-oxidation on Pt system shows spiral turbulence and global noise is introduced, this noise behaves as an overall coordinating factor that introduces a certain spatial
coherence to otherwise spatially discoordinated oscillations. In this way, chemical turbulence can be suppressed. The cases of intermittent turbulence suggest that the global
coordination is not an absolute one for intermediate noise intensities, and that the time needed for the active medium to recover plays a role in the alternation periodicity. 

\begin{figure}[h!]
\begin{center}
\includegraphics[width=13cm]{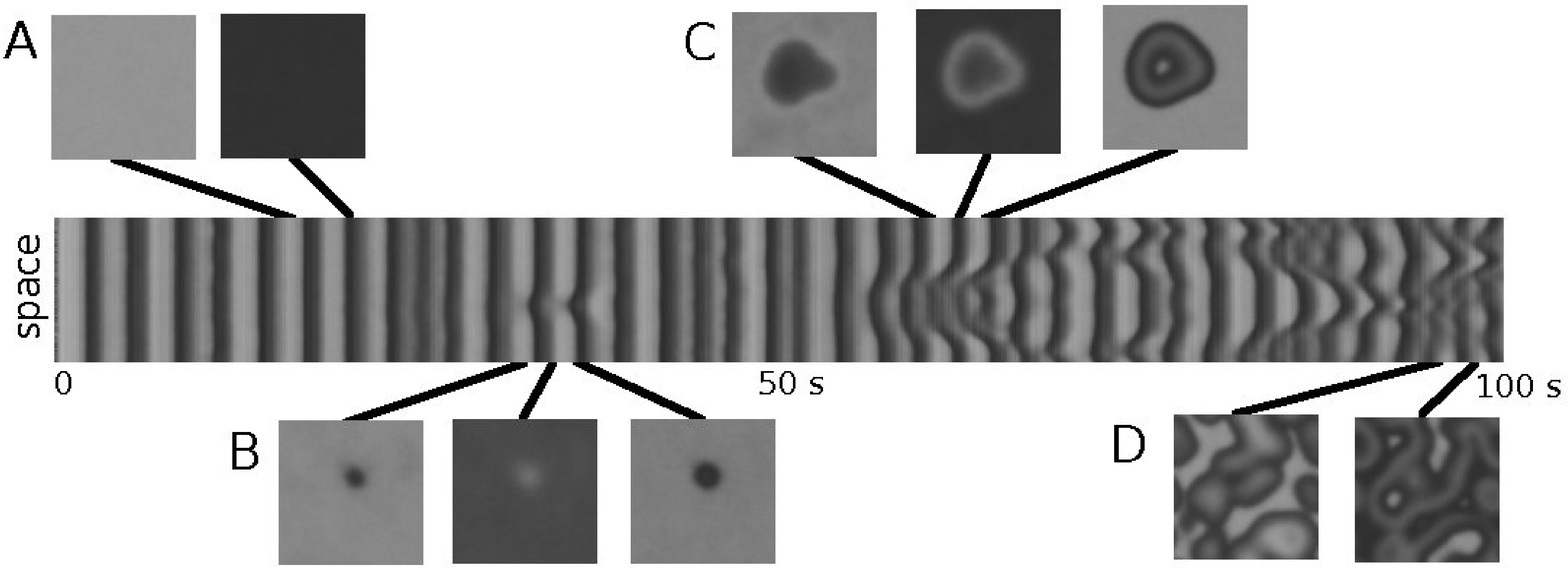}
\end{center}
\caption{Space-time plot of the evolution of the system under turbulent conditions and with external stochastic forcing. Snapshots of the system are plotted in
the pannels. Panel A shows the initial global oscillation typically observed in the numerical simulation. Panel B shows the growth of a single spot which
finally disappears. Panel C shows the generation of a pacemaker due to the intermediate amplitude of the external global noise. Panel D shows the turbulent patterns. Values
of the partial presures are:  $p_{O_{2}} = 1.3 \cdot 10^{-4}$ and $p_{CO} = 4.81 \cdot 10^{-5}$. Internal and external noises are respectively 
$\epsilon_i = 0.0005$ and $\sigma= 0.3 \% p_{CO} $. The size of the system is $0.2 \times 0.2$ $mm^2$.}
\label{f:pacemaker}
\end{figure}

In the numerical simulations performed adding external noise to the KEE
model, the effects of the fluctuations are qualitatively equivalent to the experimental case. Low noise does not modify the dynamics of the system and large noise eliminates
turbulence for similat noise intensities. For intermediate noise intensities, alternance between global oscillations and complex dynamics are also observed mimicking the
experimental results. The accuracy of the numerical simulations allows the observation of a richer phenomenology for these intermediate intensities. Moreover, other interesting
patterns induced by the noise are also witnessed in the simulations, like the formation of clusters or the generation of local pacemakers during several oscillations. Two examples
of such small pacemakers in the same numerical simulation can be seen in Fig. \ref{f:pacemaker}, where the onset of the turbulence is delayed by the noise (too weak to completely supress the
turbulence) and two transitory pacemakers appear instead of the turbulence. Similar types of patterns have been previously observed in the same system under global 
feedback \cite{bertram-mikhailov-2003}.

The control of chemical turbulence has been previously studied with other
techniques implying global resonant periodic forcing \cite{bodega_07}, feedback mechanisms \cite{bertram-beta-2003a}, or local reaction modifications via microstructures or 
laser \cite{stich_punckt_beta_08}. The forcing
amplitude, percentage of CO partial pressure, needed to effectively
suppress turbulent states was reported to be 10-20~$\%$ for the
periodically forced reaction \cite{bertram-beta-2003b}. Closer analysis of
the initial turbulent dynamics, determining the natural frequency of the
system, and performing periodical global forcing at a resonant frequency
\cite{bodega_07} allows us to reduce that needed percentage in two orders of
magnitude. As shown here, the same order of magnitude (0.1-0.2~$\%$) is enough for fast
random modulations to reach total
entrainment. It is remarkable that the latter is independent from our
previous knowledge of the system. 

The intensities of the external noise which eliminates the turbulence in the numerical simulations are also quite low (0.2-0.5 \%) and they are comparable to the case of resonant
forcing \cite{davidsen_05}.

The fluctuations also have an influence on the transition between two stable
states, anticipating a bistable transition of the system and shortening the intrinsic hysteresis loop. Using a mechanistic heuristic model, the bistable transition anticipation is
conceptually equivalent to the motion of a particle localized in the higher minimum in a tilted double-well potential \cite{Liu_Lai_Lopez_02}, in which external noise adds energy to the system to save the potential barrier. For moderate noise it has been shown that the reaction mechanism of the system more easily jumps the barrier that separates the two steady regimes.
For large noise, this barrier would become so low in front of the fluctuations that the system would always be at one of the two local minima of the potential or transiting between them.

Defect-induced fronts were experimentally observed during the anticipated bistable transitions. When the transitions were completed and the crystal surface showed a full CO
coverage, deep drops of the $p_{CO}$ signal may led to the emergence of spirals and target patterns at some defects of the surface. These oxygen structures propagated along the surface for a short
time before they disappeared.

Besides the spatiotemporal noise-induced patterns, experiments performed with noise also featured an increase in the resulting amount of CO$_{2}$ produced in the CO oxidation.
A mass spectrometer measured the CO$_{2}$ present in the UHV chamber while chemical turbulence developed, with and without modulating the $p_{CO}$. For the same amount of
reactants involved in the reaction, integrated in time, the CO$_{2}$ production rate was higher with a fluctuating CO partial pressure. 
Higher productivity thanks to fluctuating parameters is a fact already observed in other environments, namely the ozone production in atmospheres of urban areas
\cite{Liu_Lai_Lopez_02}. This fact calls attention to the possible catalytic effect of noise, which might have some impact in the increase of chemical productivity for some
other systems.

In the context of such real-world systems like marine ecosystems \cite{truscott_brindley_94,zhou_05} or ozone production \cite{Liu_Lai_Lopez_02}, in which stochasticity
processes combine with periodic annual or daily cycles, it seems worthwhile for the future to combine the two global modifications: periodic and noisy modulations of the CO oxidation reaction parameters.

In conclusion, we have shown that a stochastic external global signal introduced in the CO oxidation on
Pt(110) through the partial pressure of the CO can eliminate spatiotemporal turbulence with low
characteristic amplitudes of the noise, comparable and even smaller than in equivalent studies with the amplitudes of
periodic forcing. Furthermore, this artificially noisy signal can anticipate phase transitions in bistable conditions by reducing the reaction intrinsic hysteresis
loop.

%===================================================================\section{Acknowledgements}

\section{Acknowledgements}
\label{sect:acknowledgements}
Financial support of the EU Marie Curie Research and Training Network ``Unifying
principles in nonequilibrium pattern formation" is gratefully acknowledged. S.
A. also acknowledges support from the German Science Foundation (DFG) within the framework of Sonderforschungbereich 555 (SFB
555) ``Complex nonlinear processes".

%================================================================\clearpage

\end{document}